\pdfoutput=1

\documentclass{elsarticle}

\journal{Journal of Computational Statistics and Data Analysis}

\biboptions{authoryear, sort}

\usepackage{color}
\usepackage{amsmath}
\usepackage{amssymb}
\usepackage{bm}
\usepackage[margin=1in]{geometry}
\usepackage{amsmath}
\usepackage{bbm}
\usepackage{graphicx}
\usepackage{booktabs}
\usepackage{makecell}
\usepackage[width=.75\textwidth]{caption}
\usepackage{lineno}
\usepackage{natbib}
\usepackage{hyperref}
\usepackage{fleqn}

\usepackage{setspace} 

\usepackage{rotating}

\author[ua]{Ketong Wang}
\author[ua]{Michael D. Porter\corref{cor1}}
\address[ua]{Department of Information Systems, Statistics, and Management Science  \\ The University of Alabama, Tuscaloosa, AL 35401, United States}
\cortext[cor1]{Corresponding author. E-mail address: mdp2u@virginia.edu}

\newcommand{\thickhline}{%
	\noalign {\ifnum 0=`}\fi \hrule height 1pt
	\futurelet \reserved@a \@xhline
}

\DeclareMathOperator*{\argmin}{argmin}
\DeclareMathOperator*{\argmax}{argmax}

\begin{document}

\begin{frontmatter}
\title{Optimal Bayesian Clustering using Non-negative Matrix Factorization}

\begin{abstract}

Bayesian model-based clustering is a widely applied procedure for discovering groups of related observations in a dataset.  
These approaches use Bayesian mixture models, estimated with MCMC, which provide posterior samples of the model parameters and clustering partition. 
While inference on model parameters is well established, inference on the clustering partition is less developed. 
A new method is developed for estimating the optimal partition from the pairwise posterior similarity matrix generated by a Bayesian cluster model. 
This approach uses non-negative matrix factorization (NMF) 
to provide a low-rank approximation to the similarity matrix. 
The factorization permits hard or soft partitions and is shown to perform better than several popular alternatives under a variety of penalty functions.

\end{abstract}

\end{frontmatter}

\section{Introduction}

The goal of clustering is to discover partitions that assign 
observations into meaningful groups. 
A favorable property of Bayesian model-based clustering is that it provides versatile posterior uncertainty assessment on both the model parameters and cluster allocation estimates. 
However, while inference on model-specific parameters and mixing weights follow standard Bayesian practice, more development on estimating the clustering partition is needed. 

An intuitive, yet naive, way to obtain a point estimate of the best partition is to use a maximum a posteriori (MAP) approach which selects the partition (up to label switching) from the MCMC sample that occurs the most frequently.
But when the number of observations is large and the generating mixture is complex, the majority of the MCMC clustering samples are likely to be visited, at most, a few times. In this case, the MAP approach will not usually find the best clustering solution.

To better clarify the notion of optimal partitioning, \citet{binder1978bayesian} introduced a loss function approach. 
This considers optimal clustering as a \textit{Bayesian action} which attempts to minimize the expected loss of the partition.
Under Binder's linear loss function, the problem reduces to searching for a partition $\bm{c}^*$ that produces a binary affinity matrix $\bm{\pi}^*$ nearest to the pairwise posterior similarity matrix $\bm{\pi}$ with $\pi_{ij} = p(c_i=c_j | \bm{y})$.
The idea of using loss functions has inspired researchers to develop more advanced approaches to Bayesian clustering estimation. 
\citet{fritsch2009improved} showed that minimizing Binder's linear loss is equivalent to maximizing the Rand index. 
One extension, the adjusted Rand index, corrects for the chance of random agreement in the Rand index \citep{hubert1985comparing}.
More recently, \citet{wade2015bayesian} use a loss function based upon variation of information \citep{meilua2007comparing} as well as providing a method of assessing the uncertainty in the estimated partition, and \citet{Rastelli2017} introduce a greedy algorithm to efficiently find the partition that optimizes a large family of loss functions.  
Other influential works include \cite{quintana2003bayesian} who adapt Bayesian clustering into product partition models.
Moreover, \cite{lau2007bayesian} argue that binary integer programming is notoriously slow for Bayesian clustering estimation and provide a heuristic item-swapping algorithm that excels computationally.

We find that there are several places where these current approaches can be improved. 
The optimization strategy of directly shuffling the clustering labels to find a \emph{binary} matrix $\bm{\pi}^*$ closest to the posterior similarity  $\bm{\pi}$ can be notoriously slow.
Furthermore, conventional methods can only provide hard clustering by the nature of direct label manipulation. 
They possess no capability of accounting for ambiguous observations.
Moreover, they tend to favor two extremes which either treat all uncertain points as singleton clusters or simply lump them into a major neighbor cluster \citep{fritsch2009improved, wade2015bayesian}.

In this paper, we propose the use of non-negative matrix factorization (NMF) to identify optimal partitions from Bayesian model-based clustering models. 
We find the NMF approaches not only outperform alternative methods on clustering accuracy but can also provide deeper interpretations of the partitioning results.
Additionally, the clustering solutions produced by NMF are more compelling since they can carefully balance between the singleton-preferred and dominant-preferred extremes.

Section 2 briefly reviews the critical components of a Bayesian model-based clustering procedure and the relevant output from MCMC.
In section 3, we briefly describe the general theoretic decision framework for clustering estimation and then present the details for three current methods.
Section 4 will introduce NMF in the context of clustering estimation and further present the connections between NMF and Binder's criterion.
Simulated experiments in section 5 illustrate the clustering performance of the NMF models and their soft partitioning capabilities.
This is followed with examples from the famous galaxy and crabs datasets showing that the NMF partitioning solutions provide an attractive balance between the two clustering extremes for ambiguous data points.

\section{Bayesian Model-based Clustering}

\subsection{Bayesian Mixture Models}
Mixture models have been widely applied as a clustering tool to find meaningful patterns in a dataset. 
In Bayesian model-based clustering, observations that are determined to share the same model-specific parameters are considered to be from the same cluster.

To formally describe Bayesian model-based clustering, we introduce the notations as follows.
Let $\bm{y}$ be the observed data vector generated by a $K$-component mixture model and $\bm{c}^{\prime} = (c_1, c_2, ..., c_n)$ be the group labels corresponding to the $n$ observations.
If observation $i$ is from the $k^{th}$ component of the mixture (i.e., $c_i = k$), we say that $y_i$'s model-specific parameter is $\theta_k$ and the distribution of $y_i$ has the form $p(y_i|\theta_k)$.
Notice that each element of $\Theta = (\theta_1, \theta_2, \dots, \theta_K)$ uniquely corresponds to a cluster.
In addition, the $k^{th}$ component of the mixture has a weight or mixing probability $\phi_k$ defined as $\Pr(c_i = k)$. 

There are many approaches to estimating the model parameters, including the partition vector $\bm{c}$, in mixture models.
In frequentist approaches, such as Expectation Maximization (EM), the membership vector $\bm{c}$ is treated as a latent vector and estimated as part of the complete-data likelihood \citep{fraley2002model}.
The Bayesian alternatives usually rely on MCMC sampling to conduct estimation and posterior inference.
For example, \cite{richardson1997bayesian} use a reversible jumping MCMC algorithm which simultaneously estimates the allocation vector $\bm{c}$ as well as the number of clusters $K$.
Nonparametric Bayesian alternatives, like the Dirichlet process (DP) mixture model \citep{ferguson1973bayesian, neal2000markov}, 
consider the model parameters coming from an infinite mixture of distributions. Only a finite number of parameters will be used to model the observed (finite) data and the number of clusters $K$ from such a model is the number of unique model-specific parameters \citep{blackwell1973ferguson, rasmussen1999infinite, escobar1995bayesian, fritsch2009improved}.

\subsection{MCMC output}
A critical relationship describes the sampling process of the allocation vector $\bm{c}$:
\begin{equation}
	\label{eq:postc}
	p(\bm{c}|y) \propto \left\{\int p(y, \Psi |\bm{c})d\Psi\right\} p(\bm{c}) \,, 
\end{equation}
where $\Psi$ contains all the mixture model parameters except the allocation $\bm{c}$.
For a DP mixture model, \cite{quintana2003bayesian} write the prior distribution for the allocation vector as
\begin{equation}
	\label{eq:dpprior}
	p(\bm{c}) = \frac{\prod_{k=1}^K\alpha\Gamma(n_k)}{\prod_{i=1}^n(\alpha+i-1)} \,, 
\end{equation}
where $n_k$ is the number of observations in cluster $k$ and $\alpha$ is the DP concentration parameter. Combined with a proper setting of the mixture model \citep{dahl2005sequentially}, a Gibbs sampler can be used to obtain the $M$ MCMC samples of the partitioning vector, $\bm{c}^{(1)}, \bm{c}^{(1)}, \dots, \bm{c}^{(M)}$.
More general distributions based on DP prior can be found in \cite{ishwaran2011gibbs, pitman1997two, lijoi2007controlling}.
Additionally, \cite{lau2007bayesian} provide a detailed review of various prior distributions.

A convenient way to summarize the partition information, which is unaffected by label switching and can be used when the number of clusters $K$ is not equal for all MCMC samples, is with the pairwise posterior similarity matrix.
Intuitively, two data points are more likely to be members of the same cluster when they appear together frequently in the partitions $\bm{c}^{(1)}, \bm{c}^{(1)}, \dots, \bm{c}^{(M)}$.
To quantify this relationship, a pairwise posterior similarity matrix $\bm{\pi} = \left\{\pi_{ij}\right\}$ is defined by
\begin{equation}
\label{eq:postsim}
\pi_{ij} = p(c_i = c_j | \bm{y}) \,, 
\end{equation}
where $c_i$ and $c_j$ are the cluster assignments of observations $y_i$ and $y_j$. 
When the true probabilities are unknown, the posterior similarity can be estimated from the MCMC samples
\begin{equation}
\label{eq:postsimMCMC}
\hat{\pi}_{ij} = \hat{p}(c_i = c_j | \bm{y}) = \frac{1}{M}\sum_{m=1}^M \mathbbm{1} \{c_i^{(m)} = c_j^{(m)}\} \,, 
\end{equation}
where $\mathbbm{1} \{c_i^{(m)} = c_j^{(m)}\}$ equals $1$ if $c_i^{(m)} = c_j^{(m)}$ and $0$ otherwise.

\section{Point Estimate of Clustering}
\subsection{Expected Loss}
\label{subsec:optBayesClust}
	
Suppose a loss function $L(\bm{c}, \bm{c}^*)$ that quantifies the cost of estimating a partition $\bm{c}$ with $\bm{c}^*$. 
The posterior expected loss of using partition $\bm{c}^*$ is 
\begin{equation}
\label{eq:exploss}
E_c\left[L(\bm{c}, \bm{c}^*) | \bm{y}\right] = \sum_{\bm{c}}L(\bm{c}, \bm{c}^*)p(\bm{c}|\bm{y}) \,,  
\end{equation}
where $p(\bm{c}|\bm{y})$ is the posterior PMF of partition $\bm{c}$. 
The clustering goal is to find the Bayes optimal partition i.e., the partition which minimizes \eqref{eq:exploss}.
However, because it is often infeasible to obtain a closed-form expression of $p(\bm{c}|\bm{y})$, 
the posterior expected loss is often approximated by its sample version:
\begin{equation}
\label{eq:sampleloss}
E_{\bm{c}}\left[L(\bm{c}, \bm{c}^*) | \bm{y}\right] \approx \frac{1}{M}\sum_{m=1}^{M}L(\bm{c}^{(m)}, \bm{c}^*) \,, 
\end{equation}
where $\bm{c}^{(m)}$ is the partition from the $m^{th}$ MCMC sample.

While the approximation above is intuitive, computation can be expensive when $M$ is large.
An alternative approximation is to move the expectation inside the loss function
\begin{equation}
\label{eq:samplelossswap}
E_c\left[L(\bm{c}, \bm{c}^*) | \bm{y}\right] \approx L(E_{\bm{c}}\left[\bm{c}|\bm{y}\right], \bm{c}^*) \,, 
\end{equation}
where $E_{\bm{c}}\left[\bm{c}|\bm{y}\right]$ is the posterior expected partition. 
Note that under linear loss functions, the approximation sign in Eq. \ref{eq:samplelossswap} is a strict equality \citep{binder1978bayesian}.
Moreover, \cite{fritsch2009improved} and \cite{wade2015bayesian} argue that this approximation gives similar solutions under many useful non-linear loss functions.
To practically evaluate the approximation, one only needs to replace the posterior expectation of a label agreement $E_{\bm{c}}[\mathbbm{1}\{c_i^{(m)} = c_j^{(m)}\}]$ in the loss function by the estimated posterior similarity $\hat{\pi}_{ij}$, or that of disagreement $E_{\bm{c}}[\mathbbm{1} \{c_i^{(m)} \neq c_j^{(m)}\}]$ by $1-\hat{\pi}_{ij}$.
We will present several loss functions and illustrate how they estimate the expected loss from Eq. \eqref{eq:sampleloss} and Eq. \eqref{eq:samplelossswap}.

\subsection{Current Estimation Approaches}
\label{subsec:current}
In general, a procedure gives the point estimate of the clustering as
\begin{equation}
	\label{eq:pointest}
	\hat{\bm{c}} = \argmin_{\bm{c}^* \in \mathcal{C}}J(\bm{c}^*) \,, 
\end{equation}
where $\mathcal{C}$ is the search space and $J$ is a penalty function.
This optimization problem requires three settings: the choice of penalty function $J$, which will correspond to an estimated expected loss function, the specification of the search space $\mathcal{C}$, and the optimization algorithm to find the $\hat{\bm{c}}$ that minimizes $J$.

\subsubsection{Binder's Criterion and Variations}
An early work on Bayesian cluster analysis is \cite{binder1978bayesian}, which proposes the Binder loss function
\begin{equation}
\label{eq:binderloss}
L^{\text{\texttt{B}}}(\bm{c}, \bm{c}^*) = \sum_{i<j} l_1 \mathbbm{1}_{\{c^*_i \neq c^*_j\}} \mathbbm{1}_{\{c_i = c_j\}}  + l_2 \mathbbm{1}_{\{c^*_i = c^*_j\}} \mathbbm{1}_{\{c_i \neq c_j\}}	\,, 
\end{equation}
where $l_1$ and $l_2$ are the loss weights for two different types of mis-assignment. 
The first term represents a false negative where $i$ and $j$ actually belong to the same cluster (i.e., $c_i=c_j$) but $c^*$ disagrees and the second term represents a false positive when $i$ and $j$ are from different clusters in reality but the estimate $c^*$ assigns them to the same cluster.
When the loss weights are equal $l_1=l_2$, the expected Binder loss becomes
\begin{equation}
E_c[L^{\text{\texttt{B}}}(c, c^*) | \bm{y}] = \sum_{i<j}|\pi_{ij} - \mathbbm{1}_{\{c_i^* = c_j^*\}}| \,, 
\end{equation}
where $\pi_{ij} = E_c[\mathbbm{1}_{\{c_i = c_j\}}|y]$. 
This leads to the penalty function under Binder's loss
\begin{equation}
\label{eq:expbinderabs}
J^{\text{\texttt{B}}}(\bm{c}^*) =  \sum_{i<j}|\hat{\pi}_{ij} - \mathbbm{1}_{\{c_i^* = c_j^*\}}| \,, 
\end{equation}
which replaces $\pi_{ij}$ with the approximation $\hat{\pi}_{ij}$ from Eq. \eqref{eq:postsimMCMC}
As discussed in Sec. \ref{subsec:optBayesClust}, the linearity of Binder's loss allows the use of  Eq. \eqref{eq:samplelossswap} with equality. 

Another popular criterion for measuring the quality of a clustering solution is the expected Rand index \citep{rand1971objective}.
However, \cite{fritsch2009improved} argue that maximizing Rand index is equivalent to minimizing the Binder's loss since they have the following relationship
\begin{equation}
\label{eq:rand}
Rand(c, c^*) = 1 - \frac{L^{\text{\texttt{B}}}(c, c^*)}{\binom{n}{2}}. 
\end{equation}
Additionally, Dahl's criterion \citep{dahl2006model} uses another form of quadratic loss function 
\begin{equation}
\label{eq:dahl}
\sum_{i<j}(I_{\{c_i^*=c_j^*\}} - \pi_{ij})^2 
\end{equation}
which also can be proved equivalent to Binder's loss \citep{fritsch2009improved}. 
Thus, in this paper, we only use the original Binder's criterion to represent all the equivalent variations.

\subsubsection{PEAR Criterion}
The \emph{adjusted} Rand index adjusts for the expected number of arbitrary chance agreements allowing a more consistent comparison between clusterings with different group sizes \citep{milligan1985examination, morey1984measurement,hubert1985comparing}.
By maximizing the posterior expectation of adjusted Rand index (PEAR), \cite{fritsch2009improved} developed MaxPEAR, a partitioning method that estimates the optimal clustering from the posterior similarity matrix.

The adjusted Rand index between the candidate partitioning $\bm{c}^*$ and the true clustering $\bm{c}$ is given by
\begin{equation}
\label{eq:adjrand}
AR(\bm{c}, \bm{c}^*) = \frac{\sum_{i<j}\mathbbm{1}_{\{c_i^*=c_j^*\}}\mathbbm{1}_{\{c_i=c_j\}} - \sum_{i<j}\mathbbm{1}_{\{c_i^*=c_j^*\}}\sum_{i<j}\mathbbm{1}_{\{c_i=c_j\}}/\binom{n}{2}}{\frac{1}{2}[\sum_{i<j}\mathbbm{1}_{\{c_i^*=c_j^*\}} + \sum_{i<j}\mathbbm{1}_{\{c_i=c_j\}}] - \sum_{i<j}\mathbbm{1}_{\{c_i^*=c_j^*\}}\sum_{i<j}\mathbbm{1}_{\{c_i=c_j\}}/\binom{n}{2}} \,.
\end{equation}
Using the MCMC approximation from Eq. \eqref{eq:sampleloss}, adjusted Rand can be written as
\begin{equation}
\label{eq:expadjrand}
 \frac{1}{M}\sum_{m=1}^M AR(\bm{c}^{(m)}, \bm{c}^*) \approx E_{\bm{c}}[AR(\bm{c}, \bm{c}^*)|\bm{y}] \,.
\end{equation}
When $M$ is large, however, excessive evaluations of Eq.~\eqref{eq:adjrand} in Eq.~\eqref{eq:expadjrand} become computationally expensive.
Therefore, we use the alternative approximation implemented in \citep{fritsch2009improved}. This uses Eq. \eqref{eq:samplelossswap} to provide a solution based on the estimated posterior similarity matrix $\bm{\hat{\pi}}$, 
\begin{equation}
\label{eq:adjrandsim}
J^{\text{\texttt{PEAR}}}(\bm{c}^*) =
1 - 
\frac{\sum_{i<j}\mathbbm{1}_{\{c_i^*=c_j^*\}}\hat{\pi}_{ij}- \sum_{i<j}\mathbbm{1}_{\{c_i^*=c_j^*\}}\sum_{i<j}\hat{\pi}_{ij}/\binom{n}{2}}{\frac{1}{2}[\sum_{i<j}\mathbbm{1}_{\{c_i^*=c_j^*\}} + \sum_{i<j}\hat{\pi}_{ij}] - \sum_{i<j}\mathbbm{1}_{\{c_i^*=c_j^*\}}\sum_{i<j}\hat{\pi}_{ij}/\binom{n}{2}}. 
\end{equation}

\subsubsection{Variation of Information}
Another loss function, based on variation of information \citep{meilua2007comparing}, was recently used by \citet{wade2015bayesian} and \citet{Rastelli2017} . 
The VI loss function is formally defined by the following equation
\begin{equation}
\label{eq:vi}
L^{\text{\texttt{VI}}}(c, c^*) = \sum_{r=1}^{K}\frac{n_{r+}}{n}\log\left(\frac{n_{r+}}{n}\right) + \sum_{s=1}^{K^*}\frac{n_{+s}}{n} \log\left(\frac{n_{+s}}{n}\right)- 2\sum_{r=1}^{K}\sum_{s=1}^{K^*}\frac{n_{rs}}{n} \log\left(\frac{n_{rs}}{n}\right)\,, 
\end{equation}
where $n_{rs}$ counts the number of observations in cluster $r$ under the true clustering $\bm{c}$ while assigned to cluster $s$ by the candidate partition $\bm{c}^*$.
In addition, $n_{r+} = \sum_{s=1}^{n}n_{rs}$ and $n_{+s} = \sum_{r=1}^{n}n_{rs}$.
$K$ and $K^*$ are the number of clusters in partitions $\bm{c}$ and $\bm{c}^*$ respectively. 

We consider the penalty function introduced in \citet{wade2015bayesian} 
which approximates the expected VI loss using using Eq.~\eqref{eq:samplelossswap} 
\begin{equation}
\notag J^{\text{\texttt{VI}}}(\bm{c}^*)  = \sum_{i=1}^{n}\log\left(\sum_{j=1}^{n}\mathbbm{1}_{\{c_i^* = c_j^*\}}\right) - 2\sum_{i=1}^{n}\log\left(\sum_{j=1}^{n}\mathbbm{1}_{\{c_i^*=c_j^*\}}\hat{\pi}_{ij}\right). 
\label{eq:expvi}
\end{equation}
This is the lower bound on the expected VI loss, it avoids the computational expense of using all the MCMC sample partitions when $M$ is large and also provides accurate solutions  \citet{wade2015bayesian}.

\subsubsection{Partition Search Space}
The solution to Eq. \eqref{eq:pointest} is obtained by scanning through all possible candidate partitions $\bm{c}^*$ in the search space $\mathcal{C}$. Ideally, the search space contains all $B_n$ possible partitions, where $B_n$ is the Bell number for $n$ observations. However, this will be impractical when $n$ is large. Even the search space for a fixed $K$ number of clusters contains the Sterling number of the second kind \citep{weisstein2002stirling} 
\begin{equation}
	\left\{n \atop K \right\} = \frac{1}{K!}\sum_{j=0}^K(-1)^{K-j}{K\choose j}j^n
\end{equation}
possible partitions.
When an exhaustive search of $\mathcal{C}$ is not feasible, the search space can be reduced or a non-exhaustive search can be performed.

The most straightforward reduction method uses the MCMC clustering samples as the reduced search space.  This will provide at most $M$ unique candidate partitions \citep{dahl2006model}.
Another popular choice uses the dendrogram from applying hierarchical clustering to the pairwise dissimilarity matrix, $1-\hat{\pi}$. This method provides $n$ candidate partitions; one for every unique solution from the dendrogram.

Instead of refining the search space, other researchers attempt to resolve the optimization problem with more efficient algorithms.
For example, \cite{lau2007bayesian} propose a heuristic item-swapping algorithm aiming at outperforming the naive and slow binary integer programming algorithm.
However, \cite{fritsch2009improved} claim that the algorithm is still computationally costly for a large number of observations.
Another greedy search algorithm proposed by \cite{wade2015bayesian} aims to find a partitioning that minimizes the posterior expect loss within a neighborhood around the partition at the current iteration.
We found this algorithm to also be time-consuming in our experiments with a moderately sized dataset.
\citet{Rastelli2017} propose a greedy algorithm that, starting from an initial random partition, changes the membership of one observation at a time to the cluster that minimizes the overall loss. Multiple starts are suggested to enlarge the search space.

\section{Optimal Partitioning using NMF}
\label{sec:nfm}
Recall from \eqref{eq:pointest} that the objective of clustering is to find the partition that minimizes a specific penalty function. The existing penalty functions (\ref{eq:expbinderabs}, \ref{eq:adjrandsim}, \ref{eq:expvi}) are based on the estimated pairwise similarity matrix $\hat{\pi}$ and the binary partition matrix $\pi^* = \{\mathbbm{1}_{\{c_i^* = c_j^*\}}: i,j \in \{1,2, \ldots, n\} \}$
created from a candidate partition $\bm{c}^*$.
In essence, these methods are attempting to estimate the non-binary $\hat{\pi}$ with the binary $\pi^*$. 
This observation motivated us to consider non-binary estimators that could provide better estimates, are less computationally demanding, and provide a way to access the uncertainty in observations with ambiguous cluster assignments.

Therefore, this paper suggests a partitioning approach that uses non-negative matrix factorization (NMF) to approximate the similarity matrix $\bm{\pi}$.
Before a formal introduction of the NMF, we list some of its advantages.
First, NMF approximates the real-valued matrix $\bm{\pi}$ using a lower-rank decomposition which is also real-valued.
Without being restricted to a binary matrix, the optimal solution from a broader space could be much closer to $\bm{\pi}$. 
Second, research on NMF optimization is mature with many well-studied algorithms possessing useful properties for clustering. NMF partitioning methods also facilitate \textit{soft clustering}. 
Demonstrated by \cite{li2006relationships}, output from NMF optimization has a natural soft assignment interpretation and in fact, special NMF models are directly designed for probabilistic clustering problems \citep{ding2005equivalence, paisley2014bayesian, shashanka2008probabilistic}. 

\subsection{Formulation}
Nonnegative matrix factorization (NMF) decomposes a data matrix into lower-rank matrices which can help reveal underlying patterns of the data.
\citet{li2006relationships} have studied how various NMF constraints can help uncover different types of clustering patterns.
The connections between NMF and clustering have been well established and successfully applied to many research fields \citep{ding2013n, ding2005equivalence, KuangPD12, wang2013nonnegative, xu2003document}.

Let the posterior similarity matrix $\bm{\pi}$ be the data matrix to be approximated. In practice this would be the approximation $\hat{\bm{\pi}}$ from \eqref{eq:postsimMCMC}.
The basic NFM problem \citep{lee2001algorithms} is finding two lower-rank matrices that solve the optimization
\begin{equation}
	\label{eq:NMF}
	(\widehat{\bm{W}}, \widehat{\bm{H}}) = \argmin_{\bm{W}, \bm{H} > 0}||\bm{\pi} - \bm{W}\bm{H}||^2_F \,,
\end{equation}
where $F$ indicates the Frobenius norm, $\bm{\pi} \in \mathbb{R}^{n\times n}_+$, $\bm{W} \in \mathbb{R}^{n\times K}_+$ and $\bm{H} \in \mathbb{R}^{K\times n}_+$. 
The \textit{non-negativity} in NMF indicates that matrix elements must be non-negative. 

This least-squares type of NMF problem has a natural clustering interpretation for nonnegative data because the columns of $\widehat{\bm{W}}$ can be considered centroids 
and the columns of $\widehat{\bm{H}}$ are the cluster weights of the observations \citep{KuangPD12}.
This provides a close connection with the $K$-means algorithm \cite{kim2008sparse, ding2005equivalence}.
While $K$-means restricts $\widehat{\bm{H}}$ to be a binary matrix which implies hard assignment of cluster allocation, the NMF method results in a real-valued positive matrix $\widehat{\bm{H}}$ which contains both hard and soft clustering information.
Specifically, the hard cluster allocation of observation $i$ can be estimated using the following equation
\begin{equation}
	\label{eq:hard}
	\hat{c}_i = \argmax_{k=1:K} \widehat{H}_{ki} \,,
\end{equation}
which is similar to MAP method.
In other words, the row index of the maximum value of a column represents the cluster membership of the corresponding observation. 
In addition, the soft label assignment of $i^{\text{th}}$ observation can be obtained 
by standardizing the columns of $\widehat{\bm{H}}$
\begin{equation}
	\label{eq:soft}
	\hat{\bm{c}}_i^{\texttt{soft}} = \frac{\widehat{\bm{H}}_i}{\sum_{k=1}^K \widehat{H}_{ki}} \,,
\end{equation}
where $\widehat{\bm{H}}_i$ is the $i^{\text{th}}$ column of matrix $\widehat{\bm{H}}$. 
To interpret the soft clustering, the vector $\hat{\bm{c}}_i^{\texttt{soft}}$ describes the confidence levels of assigning observation $i$ into the clusters.
The higher the value of $\hat{\bm{c}}_{ik}^{\texttt{soft}}$, for example, the more confidence there is in assigning observation $i$ into cluster $k$.
To further understand the soft clustering nature of NMF, one can follow the claim by \cite{ding2005equivalence} that a strictly orthogonal $\bm{H}$ matrix leads to hardened partitions and a near-orthogonal $\bm{H}$ provides soft clustering which can be interpreted as posterior cluster probabilities.

\subsection{Connection to Binder's Criterion}
\label{subsub:connection}
Although non-negative matrix factorization and Binder's criterion for clustering originated in different fields, they share several similarities.
Table~\ref{tab:bindernmf} presents some connections between Binder's criterion and NMF. 

\begin{table}
	\centering
	\caption{Connections between clustering estimation using Binder's criterion and the standard least-squares NMF}
	\label{tab:bindernmf}
	\setlength{\tabcolsep}{12pt}
	\def\arraystretch{1.6}
	\begin{tabular}{lll}
		\toprule[1.3pt]
		& Binder's Criterion &  NMF Partition \\
		\hline
		optimization & $\argmin_{\bm{c}^*} \sum_{i<j}|\pi_{ij} - \pi^*_{ij}|$ & $\argmin_{\bm{W}, \bm{H}}\sum_{ij}|\pi_{ij} - \pi^*_{ij}|^2$ \\
		approximation & binary $\bm{\pi}^*$ & real $\bm{\pi}^* = \bm{W}\bm{H}$\\
		partition & hard &  hard or soft \\
		\toprule[1.3pt]
	\end{tabular}
\end{table}

The first line of Table \ref{tab:bindernmf} compares the penalty function of these two approaches.
NMF is attempting to minimize the square distance between two similarity matrices $\bm{\pi}$ and $\bm{\pi}^*$, while Binder's criterion seeks to minimize the absolute distance. But as shown in \citet{fritsch2009improved}, under Binder's criterion with binary $\bm{\pi}^*$, the absolute distance provides the same solution as the squared distance.
Regarding the approximation matrix, Binder's criterion seeks a binary matrix $\bm{\pi}^*$ that specifies a hard partition $\bm{c}^*$, while
NMF approximates $\bm{\pi}$ with a real-valued matrix $\bm{\pi}^*$ that can be decomposed into two non-negative matrices $\bm{W}$ and $\bm{H}$. These matrices can be used to obtain a hard partition \eqref{eq:hard} or soft partition \eqref{eq:soft}.

\subsection{NMF Variations and Algorithms}

Using our previous notation of penalty function, the NMF model in Eq. \eqref{eq:NMF} can be written in a general format as 
\begin{equation}
	\label{eq:NMFgeneral}
	(\widehat{\bm{W}}, \widehat{\bm{H}}) = \argmin_{\bm{W}, \bm{H} > 0} J^{\texttt{NMF}}(\bm{W}, \bm{H}) 
\end{equation}
where $J^{\texttt{NMF}}$ is a generic notation for the distance or loss between two matrices.  
While the baseline NMF model in Eq. \eqref{eq:NMF} uses the squared-error loss (NMF-ls), another popular choice of $J^{\texttt{NMF}}$ is the generalized Kullback-Leibler divergence (NMF-kl), which can capture different features \citep{brunet2004metagenes}.
In addition, two other variations of NMF models are considered. 
NMF-ns by \cite{pascual2006nonsmooth} introduces a smoothing matrix $\bm{S}$ between $\bm{W}$ and $\bm{H}$ to avoid deteriorating their sparseness.
It not only provides quality reconstruction to the original data but also produces more interpretable basis features in matrix $W$.
NMF-offset by \cite{badea2008extracting} adds an intercept term to absorb the common features of the clusters.
In this paper, we will include the four NMF models presented above.

Another important choice in NMF is the optimization algorithm.
While NMF has a variety of extensions, there are two primary types of optimization algorithms.
\citet{lee1999learning} propose multiplicative algorithms for squared-error loss (NFM-ls) and the generalized Kullback-Leibler divergence (NFM-kl) which guarantee convergence.
Another major algorithm family is based upon projected gradient descent proposed by \cite{lin2007projected} and extended by \cite{Kim15062007, kim2008nonnegative}.
For the squared-error NMF-ls model, the following two equations present the element-wise updating rules for both basis matrix $W$ and weight matrix $H$
\begin{equation}
	\label{eq:nmfW}
	H_{aj} \leftarrow H_{aj}\frac{(W^T\bm{\pi})_{aj}}{(W^TWH)_{aj}}	
\end{equation}

\begin{equation}
	\label{eq:nmfH}
	W_{ia} \leftarrow W_{ia}\frac{(\bm{\pi}H^T)_{ia}}{(WHH^T)_{ia}} \,.
\end{equation}
Algorithms for the other NMF models can be found in the original papers \citep{badea2008extracting, brunet2004metagenes, pascual2006nonsmooth}.

As pointed out by \cite{Vavasis2009}, while solving the exact NMF problem in Eq. \eqref{eq:NMF} is NP-hard, local heuristic search methods can be developed.
The basic NMF algorithms are computationally intensive; each iteration of \eqref{eq:nmfW} and \eqref{eq:nmfH} will cost at most $O(KN^2)$ operations, where $N$ is the sample size and $K$ is the number of components, but will often be much less due to the sparsity in $\hat{\pi}$. 
And while these calculations are performed for multiple initializations and for a range of $K$ values, parallel \citep{Li-Ding-2013} and GPU \citep{Mejia-Roa2015} based implementations are available. 
There have been many other developments in improving the computation of NMF, especially for large-scale problems  \citep{Kim2011fast,Wang2011Large,  Gemulla2011, Wang2016}.

The rank of the matrix factorization, $K$, also needs to be determined. 
Popular methods to select the rank $K$ include looking for an inflection point in the internal NMF loss function $J^{\texttt{NMF}}$ \citep{Hutchins2008} and cophenetic correlation \citep{brunet2004metagenes}. 
However, this paper uses the optimization functions (\ref{eq:expbinderabs}, \ref{eq:adjrandsim}, \ref{eq:expvi}) to estimate $K$. Specifically, NMF is run for a range of $K$ values and for each $K$ a hard partition is made using Eq.~\eqref{eq:hard} and then scored from one of the optimization functions. In this way, we are using NMF as a way to create and explore the search space $\mathcal{C}$.

\section{Examples}
\subsection{Synthetic Example}
\subsubsection{Clustering Performance on Gaussian Mixtures}
\label{sec:model}

\begin{table}[!b]
	\centering
	\caption{Experimental settings for the configuration triplet (\textit{separateness}, \textit{balancedness}, \textit{sphericity}) $\in$ \{\texttt{TTT}, \texttt{TTF}, \texttt{TFT}, \texttt{TFF}, \texttt{FTT}, \texttt{FTF}, \texttt{FFT}, \texttt{FFF}\}}.
\begin{tabular}{c}
$
\renewcommand{\arraystretch}{1.2}
\begin{array}{lllc}
	\toprule[1.3pt]
	& \text{\textit{\textbf{Separateness}}}  & \text{\textit{\textbf{Balancedness}}}  & \text{\textit{\textbf{Sphericity}}} \\
	\hline
	\text{\texttt{T}} & \begin{array}{l}
		\bm{\mu_1} = (3, 3)   \\
		\bm{\mu_2} = (-3, 3)  \\
		\bm{\mu_3} = (-3, -3) \\ 
		\bm{\mu_4} = (3, -3)
	\end{array} & \begin{array}{l}
		n_1 = 100   \\
		n_2 = 100  \\
		n_3 = 100 \\ 
		n_4 = 100
	\end{array} & \bm{\Sigma}_1 = \bm{\Sigma}_2 =\bm{\Sigma}_3 =\bm{\Sigma}_4 =\bm{I}_2 \\
	\hline
	\text{\texttt{F}} & \begin{array}{l}
		\bm{\mu_1} = (1.5, 1.5)   \\
		\bm{\mu_2} = (-1.5, 1.5)  \\
		\bm{\mu_3} = (-1.5, -1.5) \\ 
		\bm{\mu_4} = (1.5, -1.5)
	\end{array}& \begin{array}{l}
	n_1 = 150   \\
	n_2 = 50  \\
	n_3 = 100 \\ 
	n_4 = 30
	\end{array} & 
		\begin{array}{ll}
		\bm{\Sigma}_1 = \left[
		\renewcommand{\arraystretch}{0.7}
		\begin{array}{rr}
			2 & -0.8 \\ 
			-0.8 & 1
		\end{array}\right] &
		\bm{\Sigma}_2 = \left[
		\renewcommand{\arraystretch}{0.7}
		\begin{array}{rr}
		1 & 0.8 \\ 
		0.8 & 2
		\end{array}\right]\\[10pt]
		\bm{\Sigma}_3 = \left[
		\renewcommand{\arraystretch}{0.7}
		\begin{array}{rr}
		1 & 0.4 \\ 
		0.4 & 1
		\end{array}\right] &
		\bm{\Sigma}_4 = \left[
		\renewcommand{\arraystretch}{0.7}
		\begin{array}{rr}
		2 & 0 \\ 
		0 & 2
		\end{array}\right]
	\end{array} \\
	\toprule[1.3pt]
\end{array}
$
\end{tabular}
\label{tab:gmsetting}
\end{table}
To study the clustering performance of different partitioning models, \cite{fritsch2009improved} and \cite{wade2015bayesian} used Gaussian mixture data under various settings of \textit{separateness} and \textit{balancedness}.
However, there are many other setting parameters worthy of attention in Gaussian mixture \citep{melnykov2010finite}.
Without extending the focus into all the possible aspects of a Gaussian mixture models, we include one additional factor, $\textit{sphericity}$, which measures the roundness of a covariance matrix of the Gaussian components.

In our experiments, we simulate data from 4 components (clusters) and choose two levels, \texttt{T/F}, for each dimension of the triplet (\textit{separateness}, \textit{balancedness}, \textit{sphericity}) resulting in eight different types of datasets.
For example, \texttt{TTT} represents an easy-to-cluster dataset that is well separated, perfectly balanced, and completely spherical (i.e., identity covariance matrix).
The configuration \texttt{FFF} represents a more mixed, imbalanced, and non-spherical Gaussian mixture dataset.

\begin{figure}[!t]
	\makebox[\textwidth][c]{\includegraphics[width=1.2\textwidth]{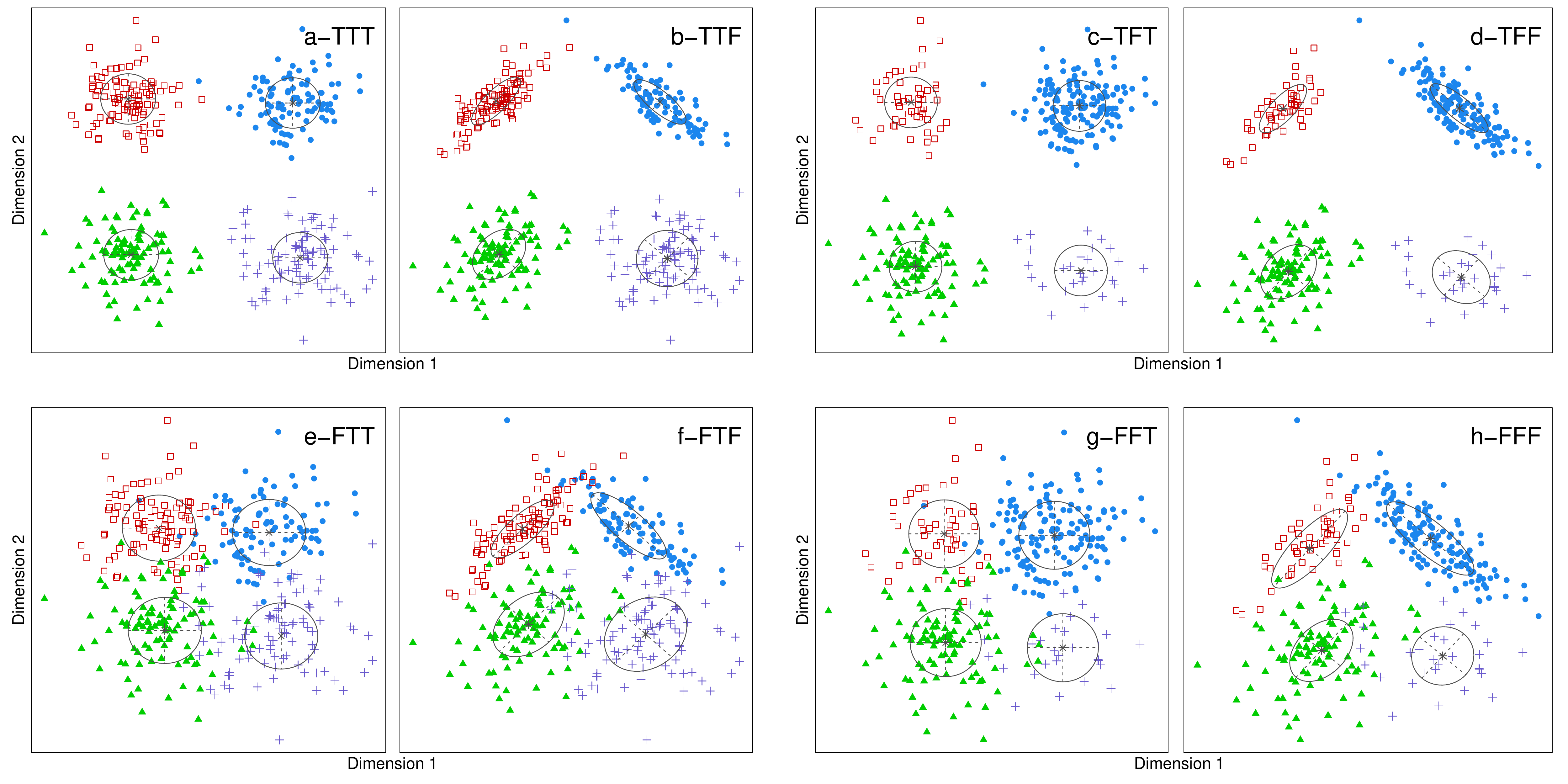}}%
	\caption{Example realizations under the eight different configurations: (\textit{separateness}, \textit{balancedness}, \textit{sphericity}). \textit{Separateness}: the top four panels $(a)-(d)$ are data with well separated Gaussian components and the bottom four $(e)-(h)$ are more mixed. \textit{Balancedness}: the left four panels $(a)(b)(e)(f)$ are data with equally sized Gaussian components and the right four panels $(c)(d)(g)(h)$ have two smaller components (i.e., the red boxes and the purple crosses). \textit{Sphericity}: within each pair of panels (e.g., the two upper-left plots $(a)(b)$), the left sub-panel $(a)$ is generated by a spherical Gaussian mixture and the right sub-panel $(b)$ has non-spherical variance-covariance matrices.}
	\label{fig:gmdata}
\end{figure}

Table \ref{tab:gmsetting} displays the values of each configuration. Specifically, the \textit{separateness} is controlled by the distance from the components means $\bm{\mu}_i$ ($i=1, 2, 3, 4$) to the origin. 
The second column shows the cluster sizes 
under the two balancedness settings and 
the last column explains the structure of the variance-covariance matrices $\bm{\Sigma}_i$.
Data is simulated 100 times for each configuration of Table~\ref{tab:gmsetting}. 
Fig.~\ref{fig:gmdata} visualizes one realization for each configuration. 

For each realization, we fit a Dirichlet process Gaussian mixture model as described in \citet{PReMiuM-package}. 
This models the data as arising from an infinite mixture of Gaussian components. The prior model for the component means are $\mu_c \sim N(\mu_0, \Sigma_0)$ and covariance matrix $\Sigma_c \sim \rm{InvWishart}(R_0, \kappa_0)$. The hyperparameters $\mu_0 = (\bar{y}_1, \bar{y}_2)$, $\Sigma_0 = diag(s_1, s_2)$ with $s_j$ the range of the $j^{th}$ covariate, $R_0$ is set to $1/2$ of the inverse of the empirical covariance matrix, and $\kappa_0=2$. These are the default and recommended values for the hyperparameters for bivariate data in the corresponding R package \texttt{PReMiuM}.

After a burn-in of 5,000, we retained 10,000 MCMC samples and created the posterior similarity matrix $\hat{\pi}$ using Eq.~\eqref{eq:postsimMCMC}. From the posterior similarity matrix the method specific partitions are obtained.
The search set, $\mathcal{C}$, for the three conventional methods is obtained from the union of the partitions generated from each unique cut of the dendrogram from agglomerative hierarchical clustering using average linkage with the dissimilarity $1-\hat{\pi}$.
The MinBinder method chooses the partition that minimizes Eq.~\eqref{eq:expbinderabs}, MaxPear chooses the partition that minimizes Eq.~\eqref{eq:adjrandsim}, and MinVI chooses the partition that minimizes Eq.~\eqref{eq:expvi}. 
We also considered choosing the partition that minimized the full loss functions (Eq.~\eqref{eq:adjrand} and Eq.~\eqref{eq:vi}), but did not find a substantial difference in performance and thus did not include the results. 
We also compared the related method of Medvedovic (Medv)
\citep{medvedovic2002bayesian, medvedovic2004bayesian} which uses complete linkage and the partition that results from cutting the dendrogram at a value close to one (we used $0.99$).

We considered four NMF models: NMF-kl, NMF-ls, NMF-ns, NMF-offset. 
For each value of $K=2,3,\ldots, 12$, the four models were fit ten times using different random initializations (all methods shared the same initializations) using the algorithms of \citet{NMF-Rpackage}. We retained the final solution that minimized the internal NMF loss (e.g., squared error). 
The optimal $K$ was then chosen by making hard partitions according to Eq.~\eqref{eq:hard} and using the partition that minimized one of the expected loss functions.

To examine how well the methods can identify the true groups, we evaluated the partitions against the three metrics: Rand index \eqref{eq:rand}, adjusted Rand (AR) index \eqref{eq:adjrand}, and the variation of information (VI) \eqref{eq:vi}.
The results are summarized in Table~\ref{tab:clustperf} which is arranged similarly to the layout  of Fig. \ref{fig:gmdata}.
Corresponding to each data panel, there is a 7-by-3 table presenting the model performances under that particular setting.
In each column under the same setting, we underline the model with the best performance.
Notice that the VI criterion is to be minimized while the Rand and AR are to be maximized. 

\begin{sidewaystable}
	\footnotesize
	\begin{center}
	\makebox[\textwidth][c]{
		\renewcommand{\arraystretch}{1.6}
		\setlength\tabcolsep{4pt}
		\begin{tabular}{lcccccccccccc}
			\toprule[2pt]
			\multicolumn{1}{c}{} & \multicolumn{3}{c}{\texttt{a-TTT}}                                                        & \multicolumn{3}{c}{\texttt{b-TTF}}                                                        & \multicolumn{3}{c}{\texttt{c-TFT}}                                                        & \multicolumn{3}{c}{\texttt{d-TFF}}                                                        \\
			\cmidrule(lr){2-4}
			\cmidrule(lr){5-7}
			\cmidrule(lr){8-10}
			\cmidrule(lr){11-13}
			& Rand                     & AR                       & VI                       & Rand                     & AR                       & VI                       & Rand                     & AR                       & VI                       & Rand                     & AR                       & VI                       \\
\hline

NMF-ls & \underline{0.967} (0.006) & \underline{0.925} (0.013) & \underline{0.167} (0.023) & \underline{0.966} (0.006) & \underline{0.921} (0.013) & \underline{0.157} (0.022) & 0.971 (0.004) & 0.938 (0.008) & 0.186 (0.018) & 0.975 (0.003) & 0.946 (0.006) & 0.164 (0.015)\\
\hline
NMF-kl & \underline{0.967} (0.006) & \underline{0.925} (0.013) & \underline{0.167} (0.023) & \underline{0.966} (0.006) & \underline{0.921} (0.013) & 0.158 (0.022) & \underline{0.985} (0.003) & 0.967 (0.007) & 0.103 (0.015) & \underline{0.997} (0.001) & \underline{0.993} (0.003) & \underline{0.029} (0.006)\\
\hline
NMF-ns & \underline{0.967} (0.006) & \underline{0.925} (0.013) & 0.168 (0.023) & \underline{0.966} (0.006) & \underline{0.921} (0.013) & \underline{0.157} (0.022) & \underline{0.985} (0.003) & \underline{0.968} (0.007) & \underline{0.100} (0.015) & 0.996 (0.001) & 0.992 (0.003) & 0.030 (0.006)\\
\hline
NMF-offset & 0.951 (0.009) & 0.894 (0.019) & 0.288 (0.047) & 0.939 (0.011) & 0.870 (0.022) & 0.329 (0.053) & 0.864 (0.005) & 0.701 (0.011) & 0.771 (0.028) & 0.858 (0.005) & 0.689 (0.012) & 0.770 (0.029)\\
\hline
MinBinder & \underline{0.967} (0.006) & \underline{0.925} (0.013) & \underline{0.167} (0.023) & \underline{0.966} (0.006) & \underline{0.921} (0.013) & 0.158 (0.022) & \underline{0.985} (0.003) & 0.967 (0.007) & 0.103 (0.015) & \underline{0.997} (0.001) & 0.992 (0.003) & 0.030 (0.006)\\
\hline
MaxPear & \underline{0.967} (0.006) & \underline{0.925} (0.013) & \underline{0.167} (0.023) & \underline{0.966} (0.006) & \underline{0.921} (0.013) & 0.158 (0.022) & \underline{0.985} (0.003) & 0.967 (0.007) & 0.103 (0.015) & \underline{0.997} (0.001) & 0.992 (0.003) & 0.030 (0.006)\\
\hline
MinVI & \underline{0.967} (0.006) & \underline{0.925} (0.013) & 0.168 (0.023) & \underline{0.966} (0.006) & \underline{0.921} (0.013) & 0.158 (0.022) & 0.984 (0.004) & 0.966 (0.007) & 0.103 (0.015) & \underline{0.997} (0.001) & 0.992 (0.003) & \underline{0.029} (0.006)\\
\hline
Medv & 0.964 (0.006) & 0.919 (0.013) & 0.181 (0.024) & 0.964 (0.006) & 0.918 (0.013) & 0.164 (0.022) & 0.984 (0.003) & 0.965 (0.007) & 0.106 (0.015) & 0.996 (0.002) & 0.991 (0.003) & 0.034 (0.007)\\

			\toprule[2pt]
			\multicolumn{1}{c}{} & \multicolumn{3}{c}{\texttt{e-FTT}}                                                        &  \multicolumn{3}{c}{\texttt{f-FTF}}                                                        & \multicolumn{3}{c}{\texttt{g-FFT}}                                                        & \multicolumn{3}{c}{\texttt{h-FFF}}                                                        \\
			\cmidrule(lr){2-4}
			\cmidrule(lr){5-7}
			\cmidrule(lr){8-10}
			\cmidrule(lr){11-13}
			& \multicolumn{1}{c}{Rand} & \multicolumn{1}{c}{AR} & \multicolumn{1}{c}{VI}   & \multicolumn{1}{c}{Rand} & \multicolumn{1}{c}{AR} & \multicolumn{1}{c}{VI}     & \multicolumn{1}{c}{Rand} & \multicolumn{1}{c}{AR} & \multicolumn{1}{c}{VI}     & \multicolumn{1}{c}{Rand} & \multicolumn{1}{c}{AR} & \multicolumn{1}{c}{VI}     \\

\hline
NMF-ls & 0.762 (0.004) & 0.466 (0.006) & 1.627 (0.012) & 0.850 (0.005) & 0.637 (0.010) & 1.243 (0.015) & 0.805 (0.005) & 0.593 (0.009) & 1.342 (0.016) & 0.876 (0.003) & 0.735 (0.007) & 0.992 (0.018)\\
\hline
NMF-kl & 0.765 (0.004) & 0.470 (0.006) & 1.630 (0.012) & 0.857 (0.005) & 0.650 (0.009) & \underline{1.216} (0.015) & 0.806 (0.005) & 0.592 (0.009) & 1.350 (0.015) & 0.882 (0.004) & 0.744 (0.007) & 0.993 (0.020)\\
\hline
NMF-ns & 0.765 (0.004) & 0.470 (0.006) & 1.623 (0.012) & 0.855 (0.005) & 0.646 (0.009) & 1.223 (0.015) & 0.805 (0.004) & 0.593 (0.008) & 1.345 (0.014) & 0.878 (0.003) & 0.738 (0.007) & 0.992 (0.018)\\
\hline
NMF-offset & 0.767 (0.005) & 0.470 (0.007) & 1.658 (0.013) & \underline{0.863} (0.005) & \underline{0.662} (0.010) & 1.229 (0.017) & 0.805 (0.005) & 0.577 (0.011) & 1.462 (0.033) & 0.869 (0.004) & 0.719 (0.008) & 1.045 (0.020)\\
\hline
MinBinder & \underline{0.778} (0.005) & \underline{0.486} (0.007) & 1.670 (0.014) & 0.859 (0.004) & 0.649 (0.009) & 1.279 (0.014) & \underline{0.819} (0.005) & \underline{0.614} (0.009) & 1.357 (0.017) & \underline{0.884} (0.003) & \underline{0.748} (0.007) & 1.006 (0.020)\\
\hline
MaxPear & 0.768 (0.005) & 0.476 (0.007) & \underline{1.616} (0.013) & 0.856 (0.005) & 0.647 (0.009) & 1.223 (0.015) & 0.813 (0.005) & 0.607 (0.009) & 1.332 (0.015) & 0.882 (0.003) & 0.745 (0.007) & 0.991 (0.020)\\
\hline
MinVI & 0.758 (0.005) & 0.463 (0.007) & 1.623 (0.012) & 0.855 (0.005) & 0.646 (0.009) & 1.221 (0.015) & 0.778 (0.005) & 0.554 (0.008) & \underline{1.323} (0.012) & 0.872 (0.004) & 0.729 (0.008) & \underline{0.987} (0.018)\\
\hline
Medv & 0.745 (0.007) & 0.445 (0.008) & 1.653 (0.012) & 0.850 (0.004) & 0.636 (0.009) & 1.230 (0.014) & 0.742 (0.013) & 0.506 (0.017) & 1.391 (0.017) & 0.873 (0.003) & 0.727 (0.007) & 1.025 (0.018)\\

			\toprule[2pt]
		\end{tabular}
	}
	\end{center}
	\caption{ Mean and standard error of the clustering performance from 100 simulations. 
		 Corresponding to the structure of Fig. \ref{fig:gmdata}, there is a 8-by-3 performance table under each of the eight settings.}
	\label{tab:clustperf}
\end{sidewaystable}

The first noteworthy result is that NMF estimation methods are consistently better than the three conventional approaches when the data are more ambiguous (i.e., \texttt{e-f-g-h}).
In other words, when the data are more mixed, the NMF-based methods can better identify the actual clustering structure. 
Even with the exceptions under setting \texttt{e-FTF}, NMF-ns and NMF-offset still provide solutions of comparable performance.
This phenomenon pertains to the fact that traditional methods tend to group uncertain points into individual clusters, which notoriously over-identifies the number of clusters.

Not surprisingly, all the estimation methods perform well on the separated, perfectly balanced, and nicely rounded data setting \texttt{a-TTT}. 
However, NMF-offset method actually outperforms other candidates under the VI criterion, which is observed in setting \texttt{b-TTF} as well.
Another observation seen by comparing \texttt{a-b} as well as \texttt{g-h} is that changing sphericity only will result in the same optimal estimation approaches.
When the simulated data are well separated and unbalanced (\texttt{c-d}), NMF-kl is the best procedure among the NFM methods and is competitive with MinVI.

\begin{figure}[!b]
	\begin{center}
		\makebox[\textwidth][c]{\includegraphics[scale=.6]{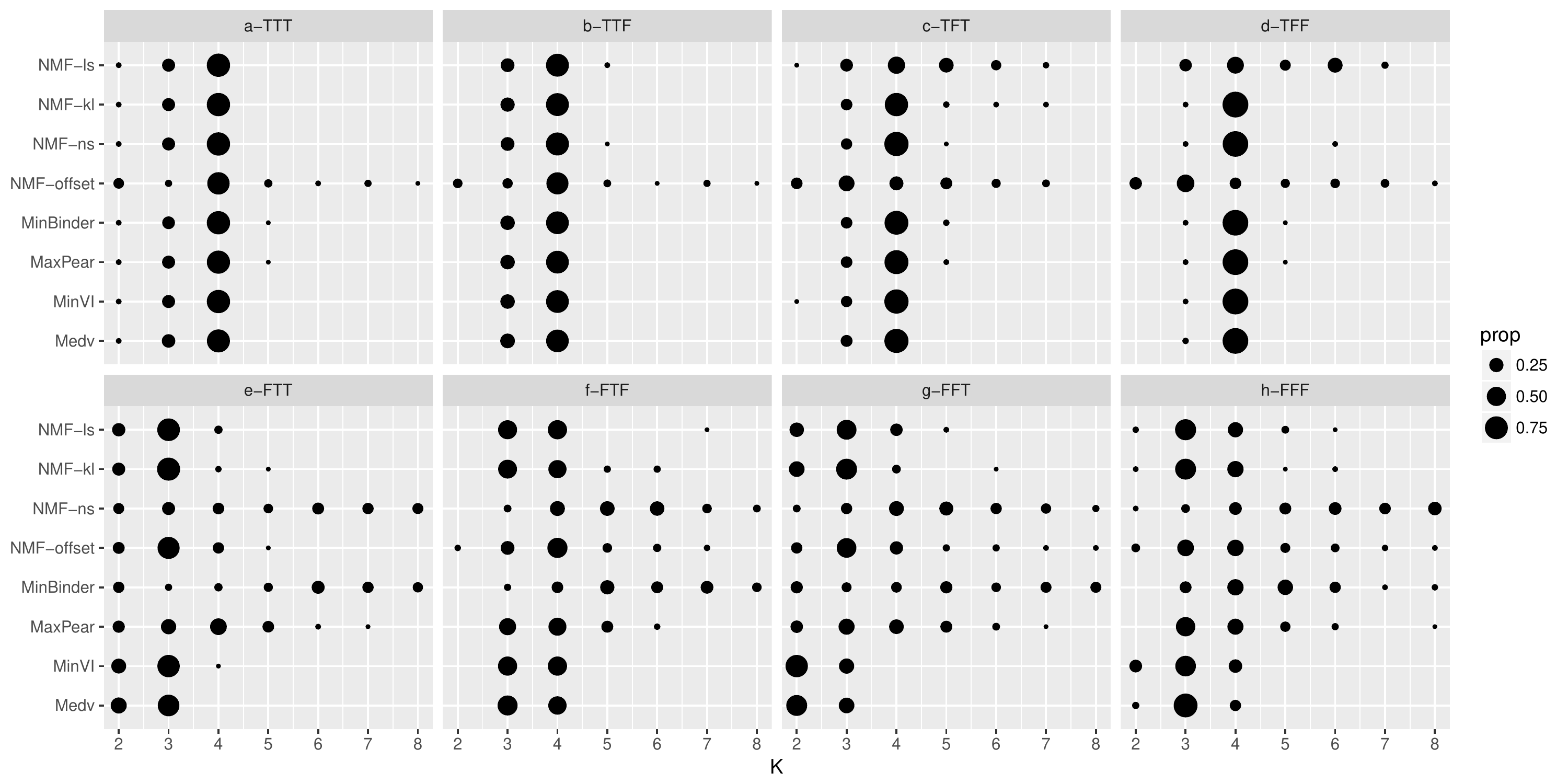}}
	\end{center}
	\caption{Distribution of the estimated number of clusters, $K$, from the 100 simulations. Values above 8 are truncated to 8.	
	}
	\label{fig:Kplot}
\end{figure}

Fig. \ref{fig:Kplot} provides a summary of the selected number of clusters $K$ for all the 100 runs.
Recall that the data were simulated with $K=4$ clusters. 
To keep perspective in the plot, $K$ is truncated at 8. 
As the cluster structure becomes more ambiguous (\texttt{e-f-g-h}), MinBinder and MaxPEAR tend to inflate the estimated cluster size. 
The MinVI and Medv approaches move in the opposite direction and tend to under-estimate the number of clusters, especially for settings $\texttt{e-FTT}$ and $\texttt{g-FFT}$.
The NMF-ls and NMF-ks provide more overall consistent estimates of the number of clusters. 
The NMF-ns does a good job of selecting the number of clusters when the clusters are well separated but suggests too many clusters when the clusters are more overlapping. The NMF-offset method does not, in general, provide a consistent estimate of the number of clusters.

\subsubsection{Soft Clustering}

\begin{figure}[!b]
	\centering
	\includegraphics[scale=.55]{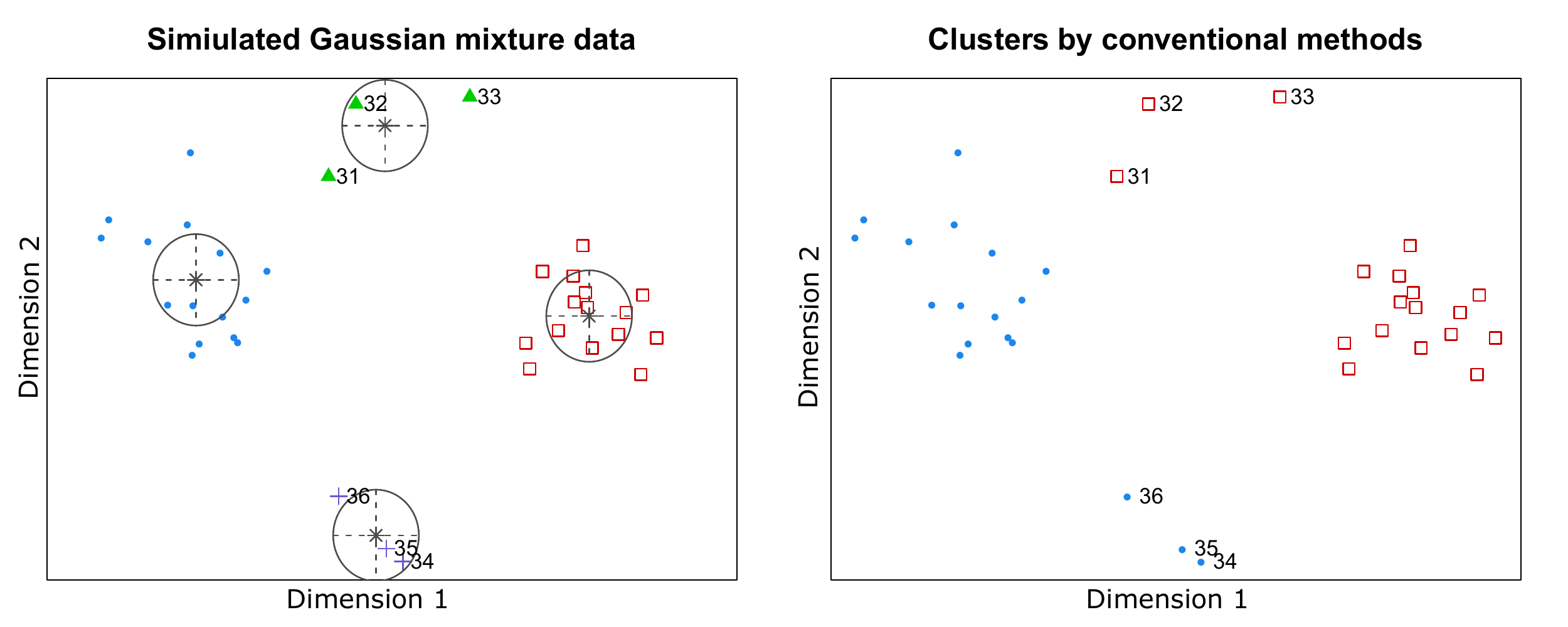}
	\caption{(Left) original data clusters with two minor groups: green-triangles (31-33) and purple-crosses (34-36). (Right) clustering solution given by the three conventional methods: MinBinder, MaxPEAR, and MinVI.}	
	\label{fig:soft_data}
\end{figure}

\begin{figure}[!b]
	\centering
	\includegraphics[scale=.35]{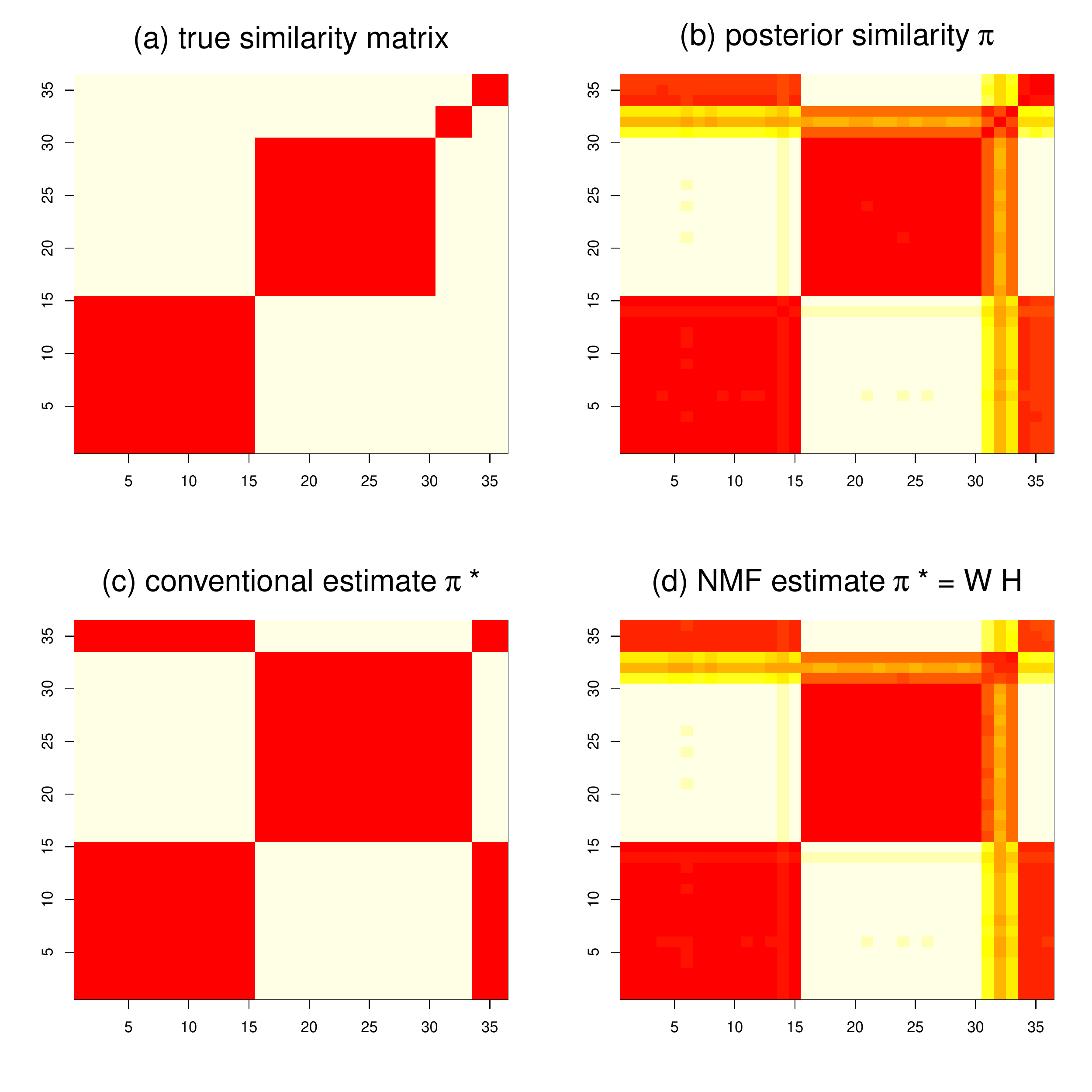}
		\caption{(a) Binary similarity matrix from the true clustering; (b) MCMC posterior similarity matrix $\bm{\pi}$ from a DP mixture model; (c) estimated similarity matrix $\bm{\pi}^*$ using conventional models; (d) estimated similarity matrix $\bm{\pi}^*$ using NMF models.}
	\label{fig:soft_heat}
\end{figure}

By providing a real-valued matrix $\bm{H}$, NMF can induce both hard partitioning labels and the soft (probabilistic) cluster assignments. 
To illustrate soft clustering, we simulated a dataset comprising four Gaussian components with two smaller clusters. 
The original data clusters are showed in the left panel of Fig. \ref{fig:soft_data}.
On the right side, we also present the hard partitioning result provided by the conventional methods (i.e., MinBinder, MaxPEAR, and MinVI).
Since they reach the same solution, only one picture is needed here.

All the conventional methods choose $K=2$ which underestimates the truth ($K=4$) by assigning observations 31-36, that form the minor clusters, to one of the larger groups.
They assign observations 31-33 to the right cluster and 34-36 to the left cluster indicating they prefer creating two elongated (non-spherical) clusters. 
While these observations do not correspond closely to the other observations in their assigned cluster, there is no way to determine this from the conventional methods. 

\begin{figure}[!t]
	\centering
	\includegraphics[scale=.6]{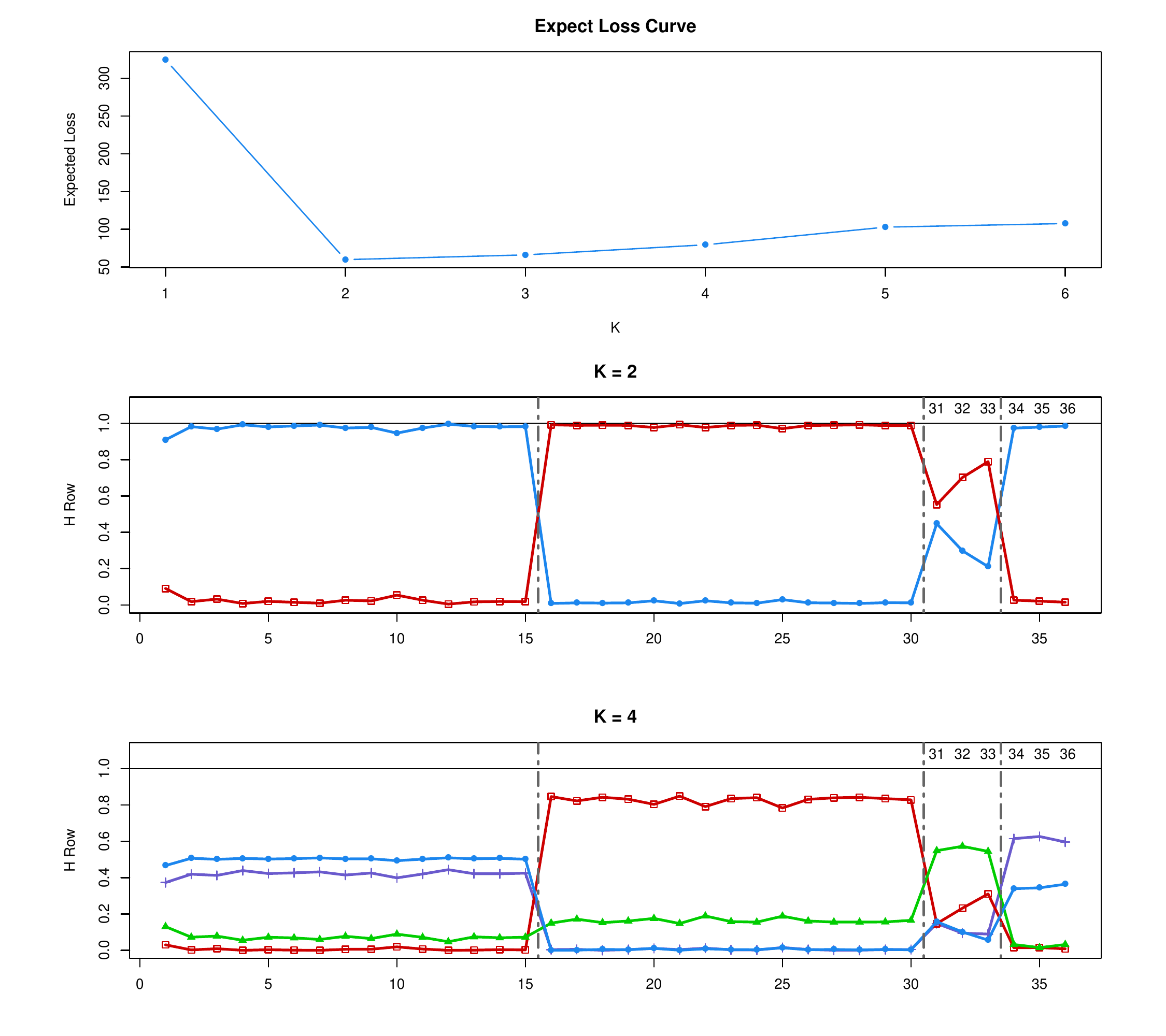}
	\caption{Soft clustering interpretation provided by the NMF-ls model. The upper panel describes the sample expected Binder's loss across various values of $K$ using the NMF estimation model. Middle and lower bottom panels visualize the rows of the $H$ matrix when $K=2$ and $K=4$ respectively.}
	\label{fig:soft_24}
\end{figure}

To illustrate the effectiveness of approximating $\bm{\pi}$ using NMF, we present the reconstructed posterior similarity matrix in Fig. \ref{fig:soft_heat} using the most generic NMF model (NMF-ls).
As clearly depicted in (\ref{fig:soft_heat}b) and (\ref{fig:soft_heat}d) the NMF solution provides a better approximation to $\bm{\pi}$ than that of the binary constrained conventional approaches showed by (\ref{fig:soft_heat}c).
The NMF solution accurately detects and restores the ambiguity around the observations from 31 to 36.
In fact, its similarity heatmap is almost identical to that of $\bm{\pi}$ in (\ref{fig:soft_heat}b).

In Fig. \ref{fig:soft_24}, we further investigate NMF's capability to identify uncertain clusters.
The top panel of Fig. \ref{fig:soft_24} shows the expected Binder's loss (Eq. \eqref{eq:expbinderabs}) under a sequence of $K$.
While $K=2$ minimizes the Binder penalty function, the loss is not much greater with $K=3$ and $K=4$ indicating that good solutions have between 2 and 4 clusters.

The middle and bottom panels of Fig. \ref{fig:soft_24} show the standardized rows (Eq. \eqref{eq:soft}) of the $\bm{H}$ matrices corresponding to the $K=2$ and $K=4$ solutions respectively. 

When $K=2$ and hard clustering is preferred, the maximum row value corresponding to each observation will determine the cluster membership obeying Eq. \eqref{eq:hard}.
For example, observations 31-33 who have larger values on the red-boxed curve will be assigned into the red cluster.
Therefore, while the hard partitioning from NMF model coincides with the ones given by conventional models, the NMF model is able to reveal the notable ambiguity on membership assignments, especially for observation 31.

In addition when $K=4$, NMF uncovers the true clustering structure where observations 31-33 and 34-36 are grouped as minor clusters.
It can also be observed that the blue-dotted and purple-crossed curves have similar profiles indicating that observations 1-15 and 34-36 have strong similarities even though they would be assigned to different clusters.
This type of soft-clustering information that is present in NMF provides a deeper understanding of the uncertainty that exists in the clustering solutions.

\subsection{Galaxy Data}

A popular dataset for research in Bayesian model-based clustering is the galaxy data \citep{roeder1990density} which records velocities of 82 galaxies from the Corona Borealis region.
While the original work aims at developing non-parametric density estimation for the one-dimension galaxy speed, latter research finds this dataset compelling for model-based clustering \citep{escobar1995bayesian, lau2007bayesian, wang2011}.

\begin{figure}[h]
	\centering
	\includegraphics[width=\textwidth]{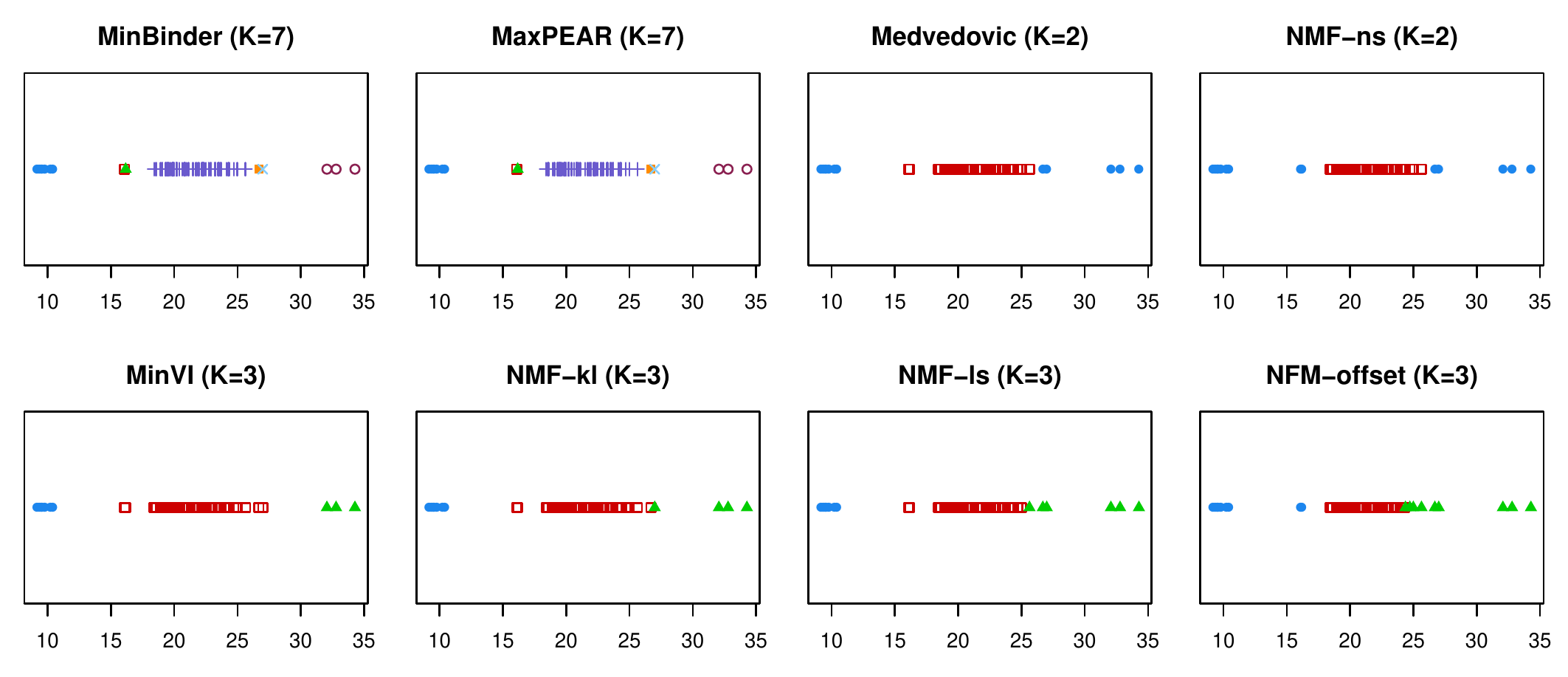}
	\caption{Clustering solutions provided by the four conventional and four NMF approaches. The $K$ in parentheses is the estimated number of components selected by the model and the symbols represent the cluster assignments. }	
	\label{fig:galaxy}
\end{figure}

In this paper, we use the MCMC samples of clusterings from a DP mixture fitted to the galaxy velocities (in 1,000 km/sec) provided by the R package \textbf{mcclust.ext} \citep{wade2015bayesian}, which also contains the predicted density using the fitted DP mixture model.
The MCMC sampling provides 10,000 draws of the clustering labels with 9,636 unique partitions (up to label switching).

Fig. \ref{fig:galaxy} shows the estimated clustering solutions for each model.
Within the traditional methods, MinBinder and MaxPEAR produce the same solution which has a relatively large $K=7$ by treating observations with high uncertainty  (i.e., observations 8, 9, 78, and 79) as singletons. 
The Medvedovic and NMF-ns methods move in the other direction and suggest there are only two clusters. Interestingly, they both indicate there is a component with large variance to account for the observations in the tails. Specifically, Medvedovic indicates observations $\{1,\ldots, 7, 78,\ldots,82\}$ are in the outlier component while NMF-ns also includes observations 8 and 9. 

The remaining approaches select $K=3$, but differ in their cluster assignments. 
The MinVI approach assigns the ambiguous observations to the dominant cluster in the middle. This suggests that each component could plausibly be  modeled with a rather symmetric distribution. 
The NMF-kl model only differs in that it assigns observation 79 into the right cluster suggesting the presence of a skewed component with a heavy tail.
The NMF-ls also includes observations 77 and 78 into the right component. 
NMF-offset is the most extreme, putting the first and last 9 components into the left and right components, respectively, suggest two highly skewed components. 

\begin{figure}[h]
	\centering
	\includegraphics[width=\textwidth]{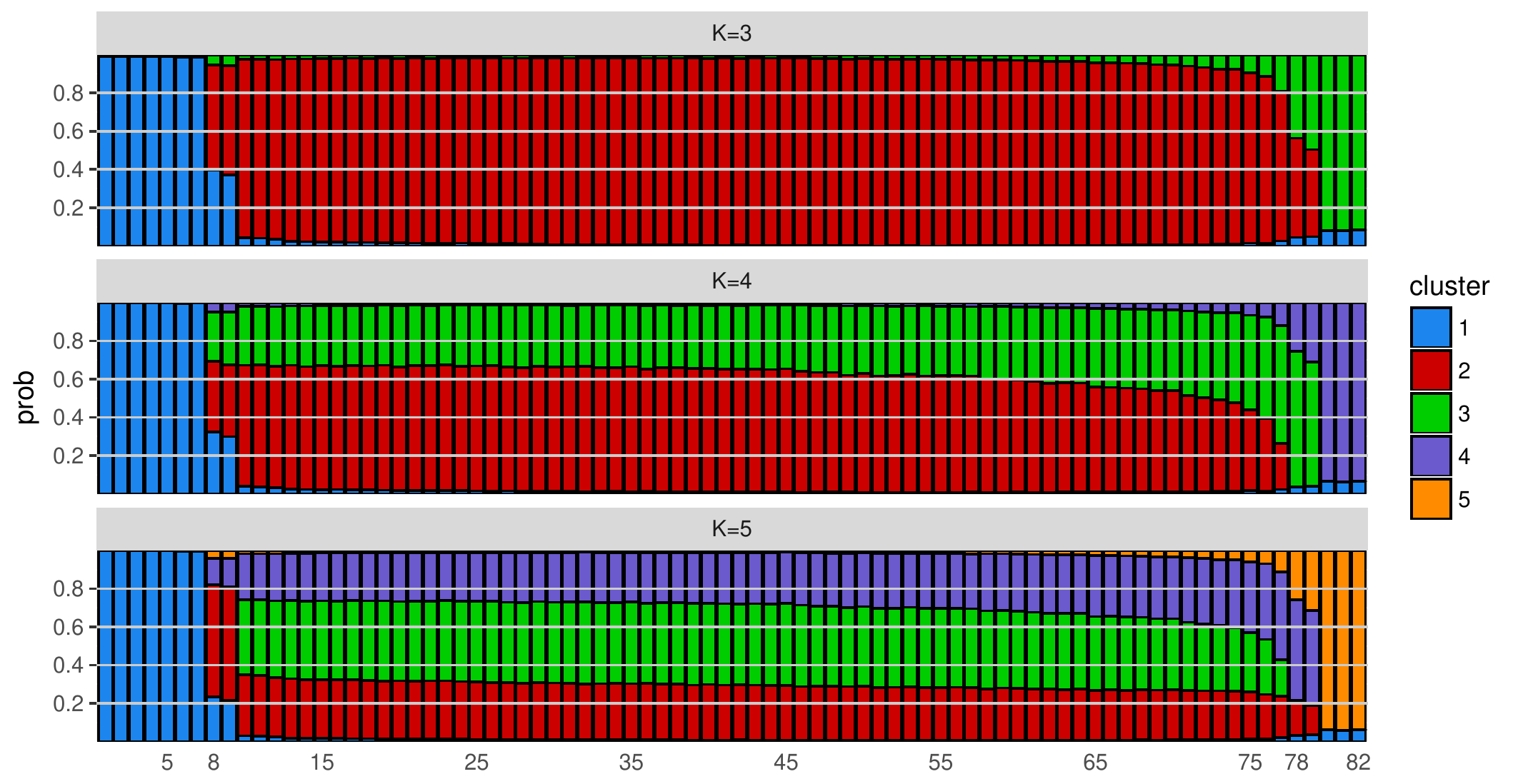}		
	\caption{Soft-clustering estimates from NMF-kl for $K=\{3,4,5\}$. The x-axis is the observation number. }	
	\label{fig:softcluster}
\end{figure}

To get a better sense of the uncertainly present in the clustering, Fig.~\ref{fig:softcluster} displays the soft-clustering estimates using NMF-kl. 
For $K=3$, this analysis reveals that the cluster assignments for observations $\{8,9\}$ and $\{78, 79\}$ are highly uncertain; something not seen from the hard classifications. 
It also shows that observation 77 has an almost 20\% chance of being part of cluster 3. 
The results for $K=4$ indicate that clusters 2 and 3 have substantial overlap but clearly prefers to include observations $\{78, 79\}$ into cluster 3. 
When $K=5$, the model is more confident about observations $\{8,9\}$ but less confident about observations $\{78, 79\}$.

\subsection{Crabs Data}

Leptograpsus crabs dataset \citep{campbell1974multivariate} has 200 crabs consisting of two species (i.e., orange and blue) each with 50 females and 50 males. This gives 4 evenly sized clusters of 50 crabs of each combination of sex and species.
Five morphological measurements are collected for each crab including front lobe size (FL), rear width (RW), carapace length (CL), carapace width (CW), and body depth (BD) all in mm.
A common, yet interesting, problem on the crabs dataset is that the number of clusters tends to be easily overestimated using all the five variables \citep{raftery2006variable}.
To overcome this problem using Gaussian mixture models,  \cite{raftery2006variable} and \cite{maugis2009variable} propose different model selection strategies to reduce variable redundancy.

\begin{figure}[!b]
	\centering
	\includegraphics[width=\textwidth]{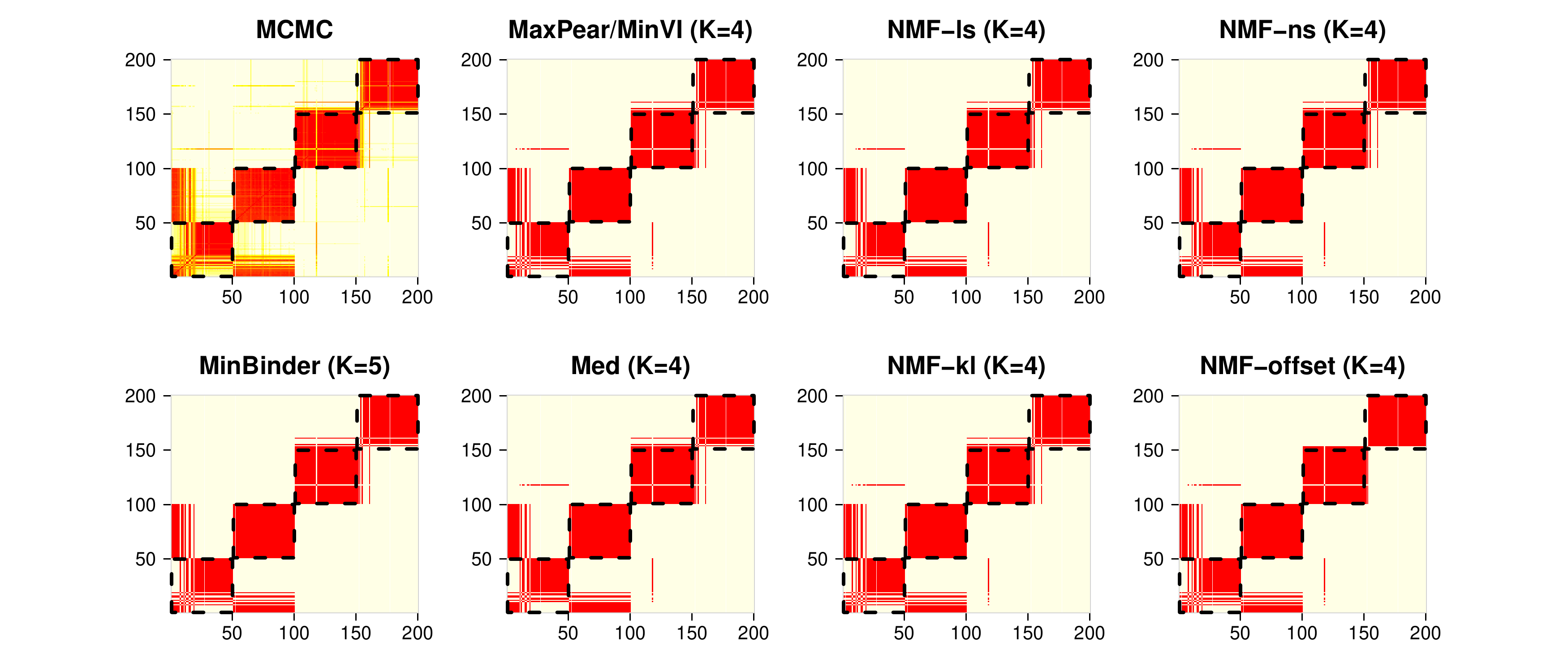} 
	\caption{Similarity and partition matrices constructed from the crabs data. Top-left is the posterior similarity matrix, $\hat{\pi}$, constructed from the MCMC samples. The other panels are the binary partition matrices, $\pi^*$ constructed from the partitioning solutions from the four conventional and four NMF methods. Dashed blocks indicate the true classifications.}
	\label{fig:crabs_combined}
\end{figure}

Fig. \ref{fig:crabs_combined} shows the posterior similarity matrix $\bm{\hat{\pi}}$ constructed from the MCMC sampling (using the same settings described in 
Section~\ref{sec:model})
Table~\ref{tab:crab_perf} shows the performance under the three clustering quality measures. 

While the posterior similarity matrix $\hat{\pi}$ reveals some uncertainty in which observations should be grouped together, all methods choose a correct number of clusters ($K=4$) except MinBinder. The clear winner in this example is the NMF-offset method, which correctly classifies the most crabs and has the best Rand, AR, and negative VI scores.

\begin{table}[!t]
	\centering
	\renewcommand{\arraystretch}{1.3}
	\caption{Clustering performance for Crabs data with known clustering labels.}
	\label{tab:crab_perf}
	\begin{tabular}{ccccccccccc}
		\toprule[2pt]
		& NMF-ls & NMF-kl & NMF-ns & NMF-offset & MinBinder & MaxPear & MinVI & Medv\\
		\hline
Rand & 0.912 & 0.915 & 0.912 & \underline{0.924} & 0.917 & 0.915 & 0.915 & 0.912\\
AR & 0.765 & 0.774 & 0.765 & \underline{0.799} & 0.779 & 0.774 & 0.774 & 0.765\\
VI & 0.762 & 0.744 & 0.762 & \underline{0.671} & 0.711 & 0.744 & 0.744 & 0.762\\	
		\toprule[2pt]
	\end{tabular}
\end{table}

\subsection{NMF Scalability}

\begin{figure}[h]
	\centering
	\includegraphics[width=.9\textwidth]{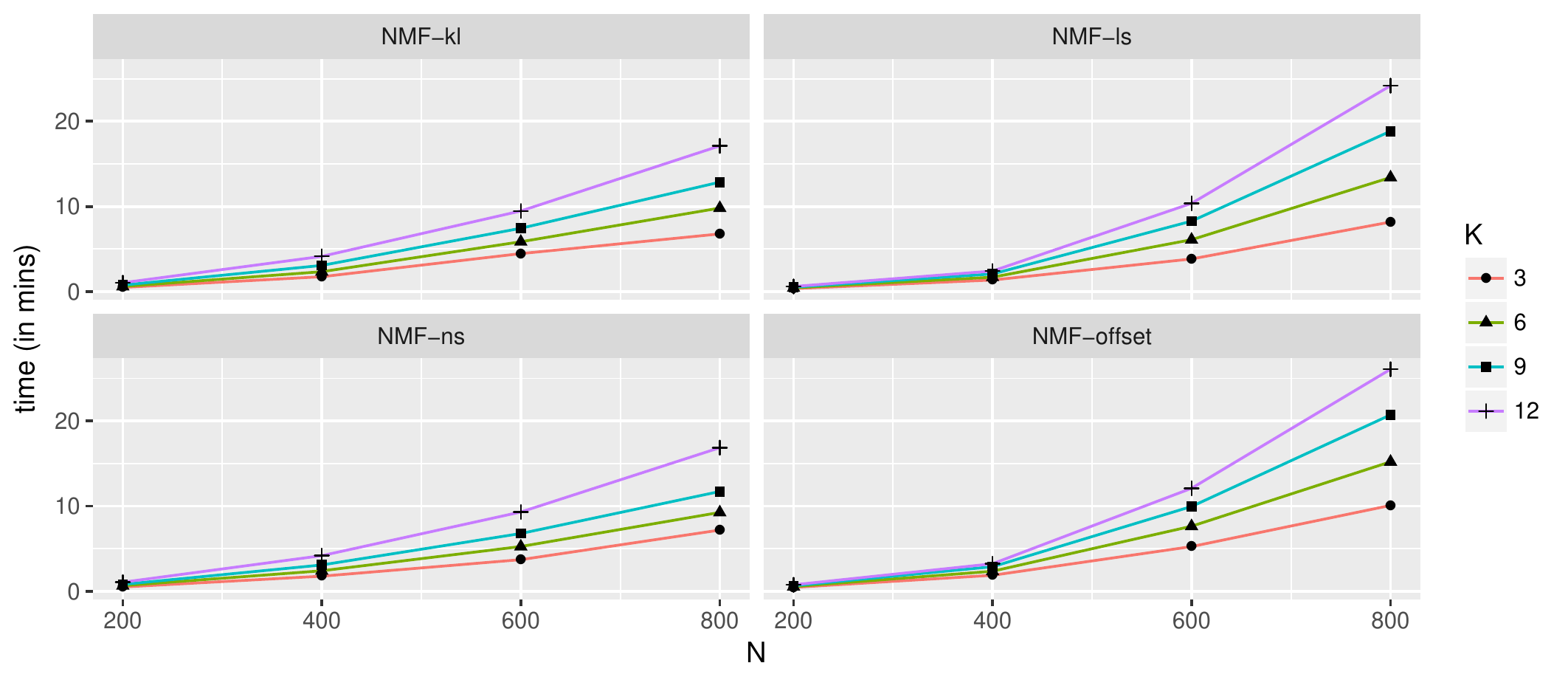}
	\caption{Computing times for the NMF methods.}	
	\label{fig:NMF-timing}
\end{figure}

Random $N \times N$ pairwise posterior similarity matrices were created for $N=200, 400, 600, 800$. 
The four NMF models were run using $K=3,6,9,12$ and the computing time is provided in  Fig.~\ref{fig:NMF-timing}. These timings are for 10 random initializations using only one core with the R package \texttt{NMF}. 
While this does show the $O(KN^2)$ scaling of NMF in a basic implementation, these times represent a rather worst case scenario as convergence will be slow (since there is no actual structure in the data) and no parallel implementation was used. 
While the \texttt{NMF} R package does facilitate parallel computation for each initialization (details given in \citet{NMF-Rvignette}), it does not include algorithms to exploit the sparseness of the posterior similarity matrices. 

\section{Conclusions}
This paper addresses the problem of estimating the optimal partition from Bayesian cluster models. 
Particularly, a non-negative matrix factorization (NMF) framework has been proposed that uses the pairwise posterior similarity matrix constructed from the MCMC clustering samples to obtain the most appropriate clustering estimates under various penalty functions.
Instead of a direct optimization using label manipulation, the NMF approach implicitly induces the partition from the weight matrix $\bm{H}$ supplied by the flexible low-rank approximation $\bm{\pi} \approx \bm{W}\bm{H}$.
Another favorable property of the NMF models is a soft/probabilistic interpretation of the clustering label assignment which can be helpful in understanding the uncertainty in the partition.

In our experiments, we found several practical advantages of the NMF partitioning models.
Their clustering capability outperformed existing approaches under various Gaussian mixture settings and performance measurements.
Additionally, NMF was able to recognize ambiguous observations located around the boundaries of major clusters.
Even under a misspecified number of clusters, NMF could still identify the uncertain data points, which is something not available with hard-clustering methods. 
The galaxy data application demonstrates that NMF can provide balanced solutions between the two extremes, favored by the conventional models, of singleton-preferred (MinBinder and MaxPEAR) and dominant-preferred (MinVI). The NMF approach can also help reveal the uncertainty in the label assignments from an analysis of soft-clustering probability estimates. 
The analysis of the crabs data further illustrates the advantages of the NMF solutions.
By providing a larger search space, the NMF approach has a better chance of identifying the optimal hard partition. This property is especially appealing when there is much uncertainly in the posterior similarity matrix. 

We have not exhausted the options available with NMF. 
There are many other divergence metrics \citep{li2006relationships}, more rigorous probabilistic soft clustering approaches \citep{zhao2015sof}, and additional models utilizing structural constraints.
With the symmetric similarity matrix $\bm{\pi}$ as an input, for example, \cite{he2011symmetric} propose a symmetric NMF model with the approximation $\bm{\pi} \approx \bm{H}^T\bm{H}$ to explicitly take advantage of the symmetry of $\bm{\pi}$.

The use of NMF for clustering from the posterior pairwise similarity matrix opens up the potential to use spectral clustering in the mixture model framework. 
\citet{ding2005equivalence}, for example, show how spectral clustering of a graph Laplacian can be written as an NMF problem. 
Building on this, if we consider the pairwise similarity matrix in \eqref{eq:postsimMCMC} to represent a graph weight matrix then methods from graph clustering, commonly called community detection, can also be used \citep{fortunato2010community}. 
This represents an exciting future direction.

The NMF approach requires optimizing over the rank $K$ of the basis matrix $\bm{W}$. This is done by running NMF for a set of $K$ values, each with multiple initializations. While this leads to an increase in computation it also provides an expanded search space $\mathcal{C}$ which gives a better chance of finding the globally optimal partition. 
The two recent methods of \citet{wade2015bayesian} and \citet{Rastelli2017} suggest greedy algorithms to judiciously explore the search space by considering a potentially different number of components at each step. 

This paper did not substantially deal with the important issue of choosing the appropriate loss function, but rather focused on demonstrating that NMF can perform well under several common loss functions. However we did find evidence in the simulation study and examples, similar to \citep{fritsch2009improved, wade2015bayesian,Rastelli2017}, that the Rand (MinBinder) and Adjusted Rand (MaxPEAR) losses tended to create singleton clusters that overestimated the number of components as well as evidence, like \citet{Rastelli2017}, that the VI (MinVI) loss can slightly underestimate the number of components in the most complex settings. Regardless of the loss function used, however, we found that an NMF method could find the correct number of components more often than the conventional approaches.

This paper develops NMF for the post-processing step of estimating the optimal partition from the pairwise posterior similarity matrix; 
we pay little attention to the choice of component model and hyperparameters that produce the similarity matrix. 
A promising future direction is the exploration of how the choice of component (e.g., symmetric, skewed, or heavy-tailed kernels) and hyperparameters interact with the aggregation used to estimate the similarity matrix to impact the optimal partitions. 

The soft-clustering interpretation of the $\widehat{\bm{H}}$ matrix \eqref{eq:soft} provides a measurement of the uncertainty in the component membership of each observation individually. 
The joint uncertainty in cluster memberships could be obtained with the credible ball method of \citet{wade2015bayesian} which requires a best hard partition and the $M$ MCMC sample partitions.

\bibliography{mybib}

\end{document}